\documentclass[aps,prd,superscriptaddress,nofootinbib,nobalancelastpage,showpacs,nobibnotes]{revtex4}
\usepackage{epsfig}

\begin{document}
\newcommand{\identity}{\:\mbox{\sf 1} \hspace{-0.37em} \mbox{\sf 1}\,}

\title
{Neutrino transport in accretion disks}

\author{R. F. Sawyer}\email{sawyer@vulcan.physics.ucsb.edu}
\affiliation{Department of Physics, University of California at
Santa Barbara, Santa Barbara, California 93106}

\begin{abstract}
We test approximate approaches to solving a neutrino transport 
problem that presents itself in the analysis of some accretion-disk 
models. Approximation \# 1 consists of replacing the full, angular-
dependent, distribution function by a two-stream simulation, 
where the streams are respectively outwardly and  inwardly directed, 
with angles $\cos \theta=\pm 1/\sqrt{3}$ to the vertical. In this 
approximation the full energy dependence of the distribution 
function is retained, as are the energy and temperature dependences 
of the scattering rates. Approximation \# 2, used in recent works on the 
subject, replaces the distribution function by an intensity function and 
the scattering rates by temperature-energy-averaged quantities. We 
compare the approximations to the results of solving the full Boltzmann 
equation. Under some interesting conditions, approximation \# 1 passes 
the test; approximation \# 2 does not. We utilize the results of our 
analysis to construct a toy model of a disc at a temperature and density 
such that relativistic particles are more abundant than nucleons, and 
dominate both the opacity and pressure. The nucleons will still provide 
most of the energy density. In the toy model we take the rate of heat 
generation (which drives the radiative transfer problem) to be 
proportional to the nucleon density. The model allows the simultaneous 
solution of the neutrino transport and hydrostatic equilibrium problems 
in a disk in which the nucleon density decreases approximately linearly 
as one moves from the median plane of the disk upwards, reaching 
zero on the upper boundary.
\end{abstract}

\maketitle

\section{Introduction}

Hyperaccreting black holes, which have been suggested as an energy source for powering gamma ray bursts, may provide
conditions in which the close-in part of an accretion disk reaches a region of density and temperature such that neutrino
radiation enters the dynamics in an important way \cite{woosley}-\cite{macfad}. In some extreme limits, the neutrino opacity of the
disk may become important, necessitating a solution to a neutrino transport problem.
A recent paper \cite{nary1} specifically addresses the issue of the effects of neutrino opacity in the innermost portion of 
such an accretion disk. Despite the transitory nature of the
phenomena, the various time scales are such that steady flow solutions of the neutrino transport problem are of interest.
Ref. \cite{nary1} uses a simplified treatment of neutrino transport in this domain, a treatment that suppresses the energy
dependence of the cross-sections in favor of  a single parameter characterizing the 
thermal averaged opacity at an average temperature of the disk, and which further eliminates
the angular degree of freedom in the Boltzmann equation in favor of a two-stream approximation
of the the angle dependence of the neutrino distribution. 

The present note investigates these two approximations in domains
characteristic of the accretion disks of ref.\cite{nary1}, by solving
the Boltzmann equation directly, where we retain a neutrino energy spectrum and full angular 
dependence of the distribution function. The two-stream approximation to this equation, where we retain the energy dependence and the local temperature dependence of the opacities, is found to work quite adequately in most of the domains that we explore. However the further simplification of ref. \cite{nary1}, in which the equation for a distribution function is replaced by an equation for intensities, the details of which are explicated in ref. \cite{nary2}, is not adequate in some domains that are of possible interest. 
We do not apply
our results to detailed physical system, but we do sketch a technique that seems useful in solving 
the kind of neutrino transport problem that arises in these systems. We also use the technique to solve a toy problem of a disk in which 
the relativistic particles dominate the neutrino opacities and the pressure, but in which nucleons still provide the energy density, and in which it is assumed that the local heat generation is proportional to the nucleon density. Hydrostatic equilibrium can be established in this case if the nucleon density decreases roughly linearly with height (i.e., the coordinate transverse to the disk) and vanishes on the disk's surface.

We consider a disk that is infinite and homogeneous in the x and y directions, with z taken as the normal 
coordinate. In appropriately scaled coordinates the surfaces of the disk are at $z= \pm 1$.
The density and temperature will in general depend on z, with $\rho (-z)=\rho(z)$ and $T(-z)=T(z)$.
 The Boltzmann equation for the neutrino distribution function, $f({\bf p},z)$,
 is now to be solved in the region $0<z<1$, with boundary conditions $f({\bf p},0)=f(-{\bf p},0)$,
and with $f({\bf p,1})=0$ for ${\bf p}\cdot \hat z <0$. There will be a prescribed amount of heat generated per unit volume at each height in the disk, $\stackrel{.}{Q}(z)$, and therefore a prescribed profile for the energy flux upward. The temperature function must be thus be adjusted so that the solution to the Boltzmann equation produces the input flux profile. 

We include both absorption and isotropic elastic scattering in the collision terms, with reaction rates, $\lambda_{a,s}$. The methods of solution could be extended easily enough to include the simple angular dependences of standard model scattering rates, but since we are trying to answer generic questions that do not depend on neutrino flavor or the neutron-proton mixture in the medium, we shall not include them. Experience with photon radiative transfer has shown that the angular dependence of Compton scattering matters little (as long at the correct weighting is included in the definition of the angular average of the scattering amplitude) in atmospheres that are (mathematically) similar to the ones that we consider \cite{angle}. Likewise, as long as there is significant absorption
in the opacities, inelasticity in scattering should not be especially consequential.

Below we exhibit first the complete Boltzmann equation for this system, followed
by two successive simplified forms.

\subsection{Boltzmann equation}
We define a variable $x=\cos(\theta)$
for the up-moving neutrinos, with $\cos(\theta)>0$ and $x=-\cos(\theta)$ for the
down-moving neutrinos. We introduce separate notations for the distribution function in the 
two hemispheres in momentum space, $f_+(z, p, x)$ for the up-moving hemisphere and
$f_-(z, p, x)$ for the down-moving hemisphere, where $p$ is the energy (or momentum) 
of the neutrino, and where we consider absorption and isotropic elastic scattering in the 
collision terms. Then we have,

\begin{eqnarray}
x{\partial f_+(z, p, x) \over \partial z}=\lambda_a[f_{eq} ( T(z),p )-f_+(z,p,x)]+
\nonumber\\
\lambda_s \Bigr[ -f_+(z,p,x)+{1 \over 2}\int_0^1 dx'\, [f_-(z,p,x')+f_+(z,p,x')] \Bigr]\, ,
\label{boltz1}
\end{eqnarray}

\rm for the up-moving stream, and 
\begin{eqnarray}
x{\partial f_-(z, p,x) \over \partial z}=-\lambda_a[f_{eq}\Bigr ( T(z),p \Bigr )-f_-(z,p,x')]-
\nonumber\\
\lambda_s \Bigr[ -f_-(z,p,x)+{1 \over 2}\int_0^1 dx'\, [f_-(z,p,x')+f_+(z,p,x')] \Bigr]\, ,
\label{boltz2}
\end{eqnarray}

for the down-moving stream. Here $f_{eq}$ denotes the Fermi function with chemical potential equal to zero,
and a temperature that depends on $z$. The rate constants, $\lambda_a$ and $\lambda_s$ 
are functions of $p$ and depend on $z$ as well, the latter through variation of the density of scatterers or through
temperature dependence of cross-sections. Of course, the temperature $T(z)$ must be adjusted so that the resulting distribution function produces the flux profile that is assumed, or calculated from other physics. \footnote{It would be of some interest to consider cases in which the ``driving chemical potential", 
$\hat \mu=\mu_e+\mu_p-\mu_n$, for electron neutrinos is not small compared to the temperature; indeed in the supernova atmosphere,
where the temperatures are a little lower than in the accretion disk, at the same value of the density, it is essential to do so. In this case the Fermi function $f_{eq}$ in (\ref{boltz1}) and (\ref{boltz2}) would need to be evaluated with chemical potential $\hat \mu$ for neutrinos and $-\hat \mu$ for anti-neutrinos. In a steady state calculation, then, $\hat \mu(z)$ and $T(z)$ (and hence the electron-proton ratio), must be self-consistently determined in the solution of the transport problem, simultaneously fitting  the requirements a of given lepton number flux (zero for example) and the externally prescribed flux profile. The only treatment in the literature that really solves this problem in the steady flow idealization is the work of Schinder and Shapiro \cite{ss} for the case of  a hot newly-formed neutron star atmosphere.\cite{ss} This work incorporates all important elastic scattering, inelastic scattering and absorption processes, as the present work does not. Its shortcomings are that it gives results for a rather specialized set of external parameters, most notably the neutrino fluxes arising from the core physics (which do not have an analogues in the disk problem), and that it provides little qualitative insight into which features of the microphysics drive which features of the results.}

\subsection{First Simplification}
We replace the continuum angular distribution by two streams at angle cos$\theta=\pm 1/ \sqrt{3}$ to the vertical.
We then have, 
\begin{equation}
{\partial f_+(z, p) \over \partial z}=\bar \lambda_a[f_{eq} \Bigr ( T(z),p \Bigr )-f_+(z,p)]+\bar \lambda_s[f_-(z,p)-f_+(z,p)]/2 \, ,
\label{cart1a}
\end{equation}

and,
\begin{equation}
{\partial f_-(z, p) \over \partial z}=-\bar \lambda_a[f_{eq}\Bigr ( T(z),p \Bigr )-f_-(z,p)]+ \bar \lambda_s  [f_-(z,p)-f_+(z,p)]/2 \, ,
\label{cart1b}
\end{equation}

where, $\bar\lambda_{a,s}= \sqrt{3} \lambda_{a,s}$, and where the local energy flux and energy densities are obtained from
\begin{equation}
{\rm Flux}(z)={1\over 4 \pi^2 \sqrt{3}} \int p^3 dp(  f_+(z, p) -f_-(z, p)) \, ,
\end{equation}

and

\begin{equation} 
\rho_{\nu} (z)= {1 \over 4 \pi^2}\int p^3 dp(  f_+(z, p) +f_-(z, p)) \, .
\end{equation}
The inside boundary condition is $f_-(0,p)=f_+(0,p)$ and the outside $f_-(1,p)$=0.

\subsection{Second Simplification}

If  in (\ref{cart1a}) and (\ref{cart1b}), the opacities were energy-independent, then we could multiply by $p^3$ and integrate, to get  equations for intensity. 
\begin{equation}
{d I_+(z) \over dz}=\bar \lambda_a[I_{eq} \Bigr ( T(z) \Bigr ) -I_+(z)]+{1 \over 2} \bar \lambda_s[I_-(z)-I_+(z)] \, ,
\label{nary1}
\end{equation}

and
\begin{equation}
{d I_-(z) \over dz}=-\bar \lambda_a[I_{eq}\Bigr ( T(z) \Bigr )-I_-(z)]+{1 \over 2} \bar \lambda_s[I_-(z)-I_+(z)] \, ,
\label{nary2}
\end{equation}
\bigskip
where $I_{eq}(T)=\int dp\, p^3 f_{eq}(T,p)$, with the inside boundary condition,  $I_-(0)=I_+(0)$, and outside condition $I_-(1)=0$.
These are the equations discussed in ref. \cite{nary2} for the case of photon radiative transfer, a solution to which, adapted to the neutrino case, is used in ref. \cite{nary1}. Of course, for the case of neutrino transport 
the rate functions are dependent on neutrino energy
as well as the temperature of the medium. Thus the set (\ref{nary1}) and (\ref{nary2}), where we are use some sort of thermal averaged  
rate functions, represents a further, and possibly drastic, approximation to the system. 

\section{Integral Equations}

We define an (absorptive) optical thickness function $\zeta(z,p)$,

\begin{equation}
\zeta(z,p)=\int_0^z \lambda_a(y,p)\, dy
\end{equation}
The Boltzmann equations for the up-moving and down-moving distributions can now be written in integral forms,
\begin{eqnarray}
f_+(z,p,x)=e^{-\zeta(z,p)/x}\Bigr \{ f_+(0,p,x)+ \int_0^z  dy x^{-1} e^{\zeta (y,p) /x}\Bigr (\lambda_a (y,p) f_{eq}(y,p,x)
\nonumber\\
\,
\nonumber\\
+\lambda_s(y,p) \Bigr [
-f_+(y,p,x)+{1 \over 2}\int_0^1 dx'[f_+(y,p,x')+f_-(y,p,x')] \Bigr ] \Bigr ) \Bigr \},
\label{boltzint1}
\end{eqnarray}
and

\begin{eqnarray}
f_-(z,p,x)=e^{\zeta(z,p)/x}\Bigr \{  \int_z^1 dy x^{-1} e^{-\zeta (y,p)/x}\Bigr (\lambda_a (y,p) f_{eq}(y,p,x)
\nonumber\\
\,
\nonumber\\
-\lambda_s(y,p) \Bigr  [
-f_-(y,p,x)+{1 \over 2}\int_0^1 dx'[f_+(y,p,x')+f_-(y,p,x')] \Bigr ] \Bigr ) \Bigr \}\,,
\label{boltzint2}
\end{eqnarray}
with  $f_+(0,p,x)=f_-(0,p,x)$ 
 supplying the boundary condition in the interior. The external boundary condition is
already incorporated into the above equations. The corresponding integral equations for the ``first simplification" 
approximation are, 
\begin{eqnarray}
f_+(z,p)=e^{-\bar \zeta (z,p)}\Bigr ( f_+(0,p)+ \int_0^z  dy  e^{\bar \zeta (y,p)}\Bigr [\bar \lambda_a (y,p) f_{eq}(y,p)
\nonumber\\
\,
\nonumber\\
+{1\over 2}\bar \lambda_s(y,p) [-f_+(y,p)+f_-(y,p)] \Bigr ] \Bigr ) \, ,
\label{cartoonint1}
\end{eqnarray}

and

\begin{eqnarray}
f_-(z,p)=e^{\bar \zeta(z,p)}\Bigr (  \int_z^1  dy  e^{-\bar \zeta (y,p)}\Bigr [\bar \lambda_a (y,p) f_{eq}(y,p)
\nonumber\\
\,
\nonumber\\
-{1 \over 2}\bar \lambda_s(y,p) [-f_+(y,p)+f_-(y,p)] \Bigr ] \Bigr )\, ,
\label{cartoonint2}
\end{eqnarray}
where  $\bar \zeta(z,p)=\int_0^z \bar  \lambda_a(y,p)\, dy$.

Beginning with an initial guess of the temperature profile and calculating
the source term $f_{eq}(z,p)$  , we perform the integral in the equation for $f_-$
 in (\ref{boltzint2}),  taking the scattering term to be zero. We then use  $f_+(0,p,x)=f_-(0,p,x)$ 
as the initial condition to calculate $f_+$ from (\ref{boltzint1}) . This zero-order result for the distribution is used
to calculate the scattering terms in (\ref{boltz1}) and (\ref{boltz2}), the result to be used in the 
next iteration. In this way, if the iteration process is convergent, we arrive at a solution corresponding
to a particular temperature distribution. Convergence problems, when scattering is large
compared to absorption, can be avoided by feeding back, in each iteration, a scattering term
that is, say,  75\% based on the distribution from the iteration one previous to the previous one.

The other part of our task is to adjust the temperature distribution so that the flux
has the $z$ dependence fixed (for now) externally. We can do this as a part of the same 
iteration procedure. After each iteration we calculate the flux from the distribution function
which we then subtract (at each point $z$) from the desired flux function. We take these numbers
multiplied by an appropriate constant and subtract them from the (numerically calculated) $(d/dz)[T(z)]$;
this difference is taken as the new $(d/dz)[T(z)]$, from which we construct the new function $T(z)$ by
integration, keeping the central temperature fixed.

In some cases there are minor manual interventions necessary to converge to a result. In particular,
for the domain in which the scattering is much larger than the absorption, it
is necessary to ramp up the strength of the scattering while iterating, at a rate that keeps the
errors in bounds at each stage, but that still gets the job done in a tolerable number of iterations.
For the opacity functions that we have taken below, and for the ``first simplification" form of the transport equation,
 it is surprisingly easy to get up to the range in which the scattering is six times the absorption, beginning in the zero'th approximation
with a monotonically decreasing temperature function. 
For the Boltzmann equation itself, with its full angular dependence,
this whole approach becomes much less efficient, owing to the $x^{-1}$ factors in the exponentials, and
we do not particularly recommend it as a technique, although we have used it for the purposes of making some of the comparisons to be summarized below.

\section{Numerical comparisons}

For the neutrino rate functions we have chosen two forms, which have the respective generic temperature, momentum, and baryon density dependence of the following cases:

(I) Opacity dominated by scattering and absorption on nucleons (under conditions of sufficiently 
high temperature so that electrons are non-degenerate and the number of neutrons is 
approximately equal to the number of protons.)

\begin{eqnarray}
\lambda_a(z,p)=\beta_a\, n_N(z)\, p^2  \, ,
\nonumber\\
\lambda_s(z,p)=\beta_s \,n_N(z) \,p^2
\nonumber\\
\label{nucopac}
\end{eqnarray}
where $n_N$ is the nucleon number density.

(II) Relativistic e$^+$, e$^-$, $\nu, \bar\nu$ atmosphere, 

\begin{eqnarray}
\lambda_a(T,p)=\alpha_a\, p \,\,T^4 \, ,
\nonumber\\
\lambda_s(T,p)=\alpha_s \,p\,\,T^4
\label{relopac}
\end{eqnarray}

First we examined a solution of the``two stream" approximation in which the temperature and momentum dependences are taken into account, that is to say, the ``first simplification" (\ref{cart1a}) and (\ref{cart1b}). We then compared with the results of solution of the full Boltzmann equations (\ref{boltz1}) and (\ref{boltz2}).  For purposes of making the comparisons we took the flux to increase linearly with $z$. Fig. 1 compares the temperature profiles for a case with fixed temperature at the outer boundary, identical flux profiles, the ``nucleonic" reaction rates given by (\ref{nucopac}), with equal scattering and absorption,  and common optical thickness (of the one-half disk) of 9.0 \footnote {We define optical thickness of the half-layer in terms of the opacity functions (\ref{nucopac}) or (\ref{relopac}) and the temperature profiles of the solutions to (\ref{cart1a}), (\ref{cart1b})
 themselves, determining a temperature-dependent mean opacity at each point, in the standard fashion, by taking the inverse of the appropriate thermal average of $\lambda_s^{-1}+\lambda_a^{-1}$, then integrating the mean opacity (evaluated at the local temperature) to obtain the optical thickness.} In this comparison and in all that follow the ordinates are plotted in arbitrary units;
the ratios of results, once the total optical thicknesses of the disk are set, are independent of the absolute values of the input temperatures and densities.

\begin {figure}[ht]
    \begin{center}
        \epsfxsize 2.75in
        \begin{tabular}{rc}
            \vbox{\hbox{
$\displaystyle{ \, { } }$
               \hskip -0.1in \null} 
} &
            \epsfbox{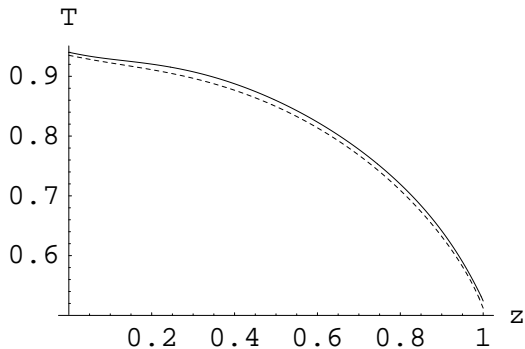} \\
            &
            \hbox{} \\
        \end{tabular}
    \end{center}
\label{fig1}
\protect\caption
    {%
Comparison of the temperature profiles in the first two-stream simplification with that obtained from solution of the full Boltzmann equation, for the case of the purely nucleonic opacity function, (\ref{nucopac}), with scattering equal to absorption. The flux function, surface temperature, and optical thickness equal to 9.0
are taken to be identical in the two cases. The flux is taken to be proportional to height, $z$. The units of temperature, $T$, are arbitrary, but the same for the two cases.  The solid curve comes from full Boltzmann equation and the dotted curve is from the two stream approximation. 
 }
\end {figure}

Fig. 2 shows the same comparison for the case of purely relativistic reaction rates given in (\ref{relopac}); the agreement is not quite as good as that shown in fig.1, but together the plots show a rather impressive agreement and provide a strong argument for using the first form of the two-stream simplification. 
\begin {figure}[ht]
    \begin{center}
        \epsfxsize 2.75in
        \begin{tabular}{rc}
            \vbox{\hbox{
$\displaystyle{ \, { } }$
               \hskip -0.1in \null} 
} &
            \epsfbox{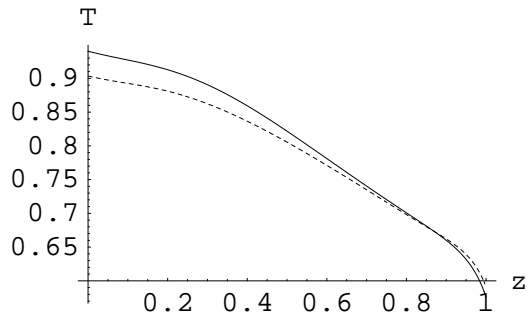} \\
            &
            \hbox{} \\
        \end{tabular}
    \end{center}
\label{fig2}
\protect\caption
    {%
The same comparison as in fig.1, now for the case of purely relativistic opacity, and optical thickness of 11.0.
 }
\end {figure}

Next we address the comparison of the``first simplification", or the two stream approximation with full temperature and energy dependence(\ref{cart1a}) and (\ref{cart1b}), and the ``second simplification" , (\ref{nary1}) and (\ref{nary2}), which subsumes all of this dependence in some average opacities. Popham and Narayan give an analytic solution for the latter equation for the case of 
$z$-independent opacities and flux that increases linearly with $z$, as in the two examples exhibited in figs. 1 and 2,

\begin{eqnarray}
(I_++I_-)/2={1+\hat \tau/2-(z^2\hat \tau )/2 \over 1+\hat \tau /2+1/ \hat \tau_a}I_e(0) \, ,
\nonumber\\
\,
\nonumber\\
(I_+-I_-)/2={z  \over 1+\hat \tau/2+1/\hat \tau _a}I_e(0)
\label{NPsln}
\end{eqnarray}

where the $z$ dependence of the source term $I_e$, given by,

\begin{equation}
 I_e(z) ={1+\hat \tau/2+1/\hat \tau _a-(z^2)( \hat \tau/2) \over 1+\hat \tau/2+1/\hat \tau_a}I_e(0)
\label{NPtemp}
\end{equation}
implicitly defines the temperature distribution.
Here $\bar \tau$ and $\bar \tau_a$ are the respective total and absorption depths of the half-layer (times the
factor $\sqrt{3}$ that enters the definition of the $\bar \lambda$ functions, above). 
In fig. 3 we show a comparison of this solution with that of the ``first simplified form" for the case of pure nucleonic opacity, with  optical thickness of $11.0$
and the same flux profile, this time displaying the neutrino energy densities $w(z)$ as a function of height, in the two cases. 
The ``second simplification'' approximation is seen to work quite well,
by the measure of the ``first simplification" approximation, and therefore, in view of the previous comparisons, as a simulation of the full transport problem.
\begin {figure}[ht]
    \begin{center}
        \epsfxsize 2.75in
        \begin{tabular}{rc}
            \vbox{\hbox{
$\displaystyle{ \, { } }$
               \hskip -0.1in \null} 
} &
            \epsfbox{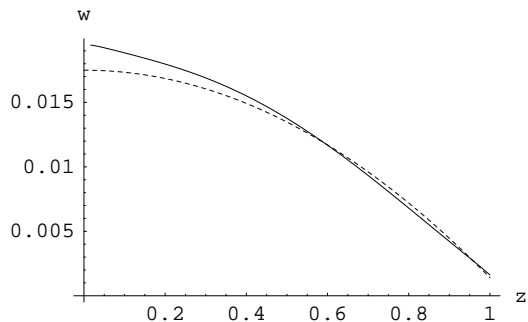} \\
            &
            \hbox{} \\
        \end{tabular}
    \end{center}
\label{fig3}
\protect\caption
    {%
The solid curve shows the neutrino energy density $w$ as a function of height, as determined by the ``first simplification" form of the equations, for the case of nucleonic opacity with equal elastic scattering and absorption rates, and optical thickness of  11.0, as in fig.1. The dotted curve shows the same plot for the solution (\ref{NPsln}) given Popham and Narayan and specialized to the same parameters.
 }
\end {figure}

The case of purely relativistic opacity is more problematic. In fig 4. we show neutrino energy density plots, again for the case of equal scattering and absorption. The ``second simplification" approximation here gives a greater than 30\% overestimate of the (central energy density)/(flux) ratio. But more significant, perhaps, are the shapes of the curves; with the same flux profile, the ``first simplification" approximation has a quite different energy density profile from the ``second simplification" result. In particular, note that the negative of the gradient of the energy density, in the outer region, is a decreasing function of $z$ in the entire region $z>.5$ in our treatment that takes into account the energy and temperature dependences, while it is positive throughout in the case in which averaged opacities are used.

\begin {figure}[ht]
    \begin{center}
        \epsfxsize 2.75in
        \begin{tabular}{rc}
            \vbox{\hbox{
$\displaystyle{ \, { } }$
               \hskip -0.1in \null} 
} &
            \epsfbox{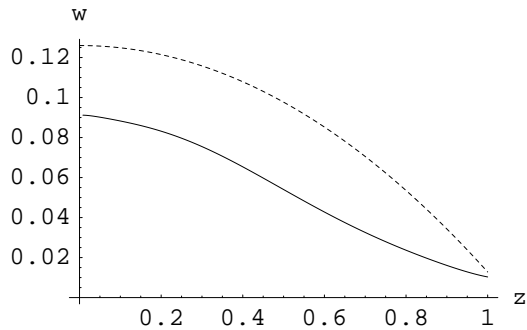} \\
            &
            \hbox{} \\
        \end{tabular}
    \end{center}
\label{fig4}
\protect\caption
    {%
The same as fig. 3, but for purely relativistic opacity, with optical thickness 11.0 as in fig.1. 
}
\end {figure}

When we plot the $T(z)^4$ instead of the energy density of the neutrinos we find a shape that differs appreciably from that shown in fig. 4 only in a region extremely near the surface (in fact, a region beginning above the average height for the last neutrino scattering). Thus the gradient of the pressure of all of the relativistic particles, taken together, declines steadily as we move upwards in the upper parts of the disk. Hydrostatic equilibrium requires that the total pressure
gradient be proportional to $z \rho (z)$, where $\rho (z)$ is the total energy density, since the operative tidal gravitational field
in the disk is proportional to $z$. Therefore if relativistic pressure were the whole story, the density would have to be a fairly rapidly decreasing function of $z$, in contrast to the uniform density assumption that appears to be used in \cite{nary1}. In any of the regimes that we here consider, the total energy density is dominated by nucleons. But both pressure and neutrino opacities are dominated by relativistic particles when the temperature is such that the ratio of the number of relativistic particles to the number of nucleons is large.

One can therefore find regimes in which the relativistic particles predominate in both the pressure and opacity calculations while the mass density is still dominated by nucleons. We might achieve hydrostatic equilibrium in such a system by having a nucleon density that decreases significantly from the center to the top. For example, consider a nucleon density that is proportional to $(1-|z|)$. Now assume as well that heat production is proportional to this density. Then we need to solve the neutrino transport problem, taking a flux profile proportional to $z-z^2/2$ (for $z>0$). The resulting relativistic pressure profile is shown in fig. 5 for the case of (optical thickness=11.0). \footnote {The total relativistic pressure is the sum of the neutrino pressure and that of the e$^\pm$ particles and the photons, the latter two computed from the local temperature; for the neutrino contribution we take the local energy density, divided by 3, from the solution of the radiative transfer problem. The shape of the energy density curve is the same as that of the total relativistic pressure except in the very outermost region.}

\begin {figure}[ht]
    \begin{center}
        \epsfxsize 2.75in
        \begin{tabular}{rc}
            \vbox{\hbox{
$\displaystyle{ \, { } }$
               \hskip -0.1in \null} 
} &
            \epsfbox{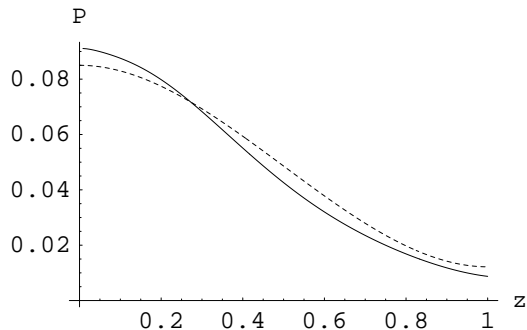} \\
            &
            \hbox{} \\
        \end{tabular}
    \end{center}
\label{fig6}
\protect\caption
    {%
The solid curve shows the total pressure (in arbitrary units) of relativistic particles as a function of height, as determined by the ``first simplification" form of the equations, for the case of relativistic opacity with equal elastic scattering and absorption rates, where the flux is now taken as proportional to $z-z^2/2$. The optical thickness is $11.$ The dotted curve shows the function, $a+b(z^2/2-z^3/3)$ that most nearly fits the pressure curve
 }
\end {figure}

On the same curve we have drawn the profile of the pressure required for hydrostatic equilibrium for the case in which the total energy density is proportional to $(1-|z|)$, giving $P={\rm const}\times (z^2/2-|z|^3/3)$. It is clear that with a small redistribution of the
nucleons, in this model we could simultaneously achieve our three goals: (a) Solution of the radiative transfer problem;
(b) hydrostatic equilibrium; and (c) matching the energy flux to a local heating rate that is proportional to the nucleon density.

In the case of accretion disks that are more or less uniform in density it appears that the temperature will never be quite high enough, at a given nucleon density, to lead to the complete dominance of pressure and opacity by the relativistic part of the medium 
\cite{nary1}. But the above example does suggest the possibility of stable relativistic outlying regions, still well under the height of (average) last neutrino scattering, in which nucleon densities decrease markedly with height. The surface of the
disk cannot be as simple as in this idealized model and one can easily imagine instabilities that lead to
escape of some of the surface plasma. Thus it might be possible to achieve an electron-positron
plasma in the region both near the surface and the axis that is more dense than that coming from neutrino-
neutrino collisions in the vicinity of the axis, the latter a source that has been discussed in the literature \cite{janka}.

The author is indebted to Andrei Beloborodov and to Chris Fryer for interesting conversations.


\begin{thebibliography}{99}
\bibitem{woosley} S. E. Woosley, Astrophys. J., {\bf 405}, 273 (1993)
\bibitem{janka}M. Ruffert, M; , H-Th Janka; K. Takahashi; and G. Schafer, Astron. Astrophys. {\bf 319}, 122 (1997)
\bibitem{pwf} R. Popham, S. E. Woosley, and C. Fryer, Astrophys. J. {\bf 356}, 356 (1999)
\bibitem{macfad} A. I. MacFadyen and S. E. Woosley, Astrophys. J.  {\bf 524}, 262 (1999)
\bibitem{nary1}T. DiMatteo, R. Perna, and R. Narayan, Astrophys. J. {\bf 579}, 706 (2002)
\bibitem{nary2}R. Popham, and R. Narayan, Astrophys. J. {\bf 442}, 337 (1995)
The relevant formulae in this paper were derived by simplifying a more exact result contained in 
a paper by I. Hubeny, Astrophys. J. {\bf 351},632 (1990).
\bibitem{angle}D. Mihalas, {\it Stellar Atmospheres}( Freeman, San Francisco, 1978)   
\bibitem{ss}P. J. Schinder and S.L. Shapiro, Astrophys. J. {\bf 259}, 311 (1982)
\end{thebibliography}
\end{document}